\documentclass[letterpaper, 10 pt, conference]{ieeeconf}  

\IEEEoverridecommandlockouts                              

\newif\ifmargincomments 
\margincommentstrue

\ifmargincomments

\else

\fi

\usepackage{amsmath,amssymb,amsfonts}
\usepackage{algorithmic}
\usepackage{graphicx}
\usepackage{textcomp}

\usepackage[utf8]{inputenc}

\usepackage[usenames,dvipsnames,svgnames]{xcolor}

\usepackage{todonotes}

\usepackage[acronyms,nogroupskip,nopostdot,nonumberlist,order=word]{glossaries}
\usepackage{glossary-mcols}

\usepackage[backend=biber, natbib=true, style=ieee,doi=false,maxbibnames=5]{biblatex}

\usepackage{bm}

\usepackage{bbold}

\usepackage{accents}

\usepackage{comment}

\usepackage[nocomma,c1]{optidef}

\usepackage{caption,subcaption}
\usepackage{units}

\usepackage[autolanguage]{numprint}

\let\labelindent\relax
\usepackage{enumitem}
\setitemize{noitemsep,topsep=0pt}

\usepackage{siunitx}

\usepackage{setspace}

\usepackage[hidelinks]{hyperref}
\usepackage[capitalize]{cleveref}

\usepackage{tabularx}

\AtEveryBibitem{\clearfield{pages}}

\glsdisablehyper 

\newcommand{\T}{^T}

\DeclareMathOperator{\reals}{\mathbb{R}}

\DeclarePairedDelimiter{\parentheses}{(}{)}

\DeclarePairedDelimiter\myabs{\lvert}{\rvert}%
\DeclarePairedDelimiter\mynorm{\lVert}{\rVert}%

\makeatletter
\let\oldmyabs\myabs
\def\myabs{\@ifstar{\oldmyabs}{\oldmyabs*}}
\let\oldmynorm\mynorm
\def\mynorm{\@ifstar{\oldmynorm}{\oldmynorm*}}
\makeatother

\newcommand{\csum}[1]{\sum_{\scriptstyle\mathclap{#1}}}

\crefname{appsec}{Appendix}{Appendices}

\newcommand{\jvspace}[1]{}

\newglossary[tlg_names]{names}{tld_names}{tdn_names}{names}
\newglossary[tlg_amod]{amod}{tld_amod}{tdn_amod}{AMoD system}

\makeglossaries

\setacronymstyle{long-short}
\newacronym{opf}{OPF}{Optimal Power Flow}
\newacronym{lp}{LP}{linear program}
\newacronym{milp}{MILP}{mixed-integer linear program}
\newacronym{amod}{AMoD}{Autonomous Mobility on Demand}
\newacronym{eamod}{E-AMoD}{Electric Autonomous Mobility-on-Demand}
\newacronym{pamod}{P-AMoD}{Power-in-the-Loop Autonomous Mobility-on-Demand}
\newacronym{ev}{EV}{electric vehicle}
\newacronym{av}{AV}{autonomous vehicle}
\newacronym{pge}{PG\&E}{Pacific Gas and Electric Company}
\newacronym{iso}{ISO}{independent system operator}
\newacronym{v2g}{V2G}{vehicle-to-grid}
\newacronym{soc}{SoC}{state-of-charge}
\newacronym{pdn}{PDN}{power distribution network}
\newacronym{taz}{TAZ}{transportation analysis zones}
\newacronym{aaa}{AAA}{American Automobile Association}
\newacronym{tou}{TOU}{time-of-use}
\newacronym{afdc}{AFDC}{Alternative Fuels Data Center}
\newacronym{vmt}{VMT}{vehicle miles travelled}
\newacronym{epa}{EPA}{Environmental Protection Agency}
\newacronym{wltp}{WLTP}{World harmonized Light-duty vehicles Test Procedure}
\newglossaryentry{leaf}
{
		name={Leaf S},
	description={2020 Nissan Leaf S}
}

\newglossaryentry{spring}
{
		name={Spring},
	description={2021 Dacia Spring}
}

\newglossaryentry{model3}
{
		name={Model 3},
	description={2020 Tesla Model 3 Long Range AWD}
}
\newglossaryentry{indicator}
{
	name={\ensuremath{\mathbb{1}}},
	description={indicator function}
}

\newglossaryentry{t}
{
	name={\ensuremath{t}},
	sort={t},
	description={time}
}

\newglossaryentry{t_set}
{
	type=amod,
	name={\ensuremath{\mathcal{T}}},
	sort={T},
	description={set of time steps}
}

\newglossaryentry{n_t}
{
	name={\ensuremath{T}},
	sort={T},
	description={number of time steps}
}

\newglossaryentry{t_step}
{
	type=amod,
	name={\ensuremath{\Delta\gls{t}}},
	sort={zz D t},
	description={length of a time step}
}

\newglossaryentry{l}
{
	name={\ensuremath{l}},
	sort={l},
	description={location}
}

\newglossaryentry{l_set}
{
	type=amod,
	name={\ensuremath{\mathcal{L}}},
	sort={L},
	description={set of locations}
}

\newglossaryentry{n_l}{
    name={\ensuremath{L}},
    sort={L},
    description={number of locations}
}

\newglossaryentry{c}
{
	name={\ensuremath{c}},
	sort={c},
	description={charge level}
}

\newglossaryentry{c_set}
{
	type=amod,
	name={\ensuremath{\mathcal{C}}},
	sort={C},
	description={set of charge levels}
}

\newglossaryentry{n_c}{
    name={\ensuremath{C}},
    sort={C},
    description={number of charge levels}
}

\newglossaryentry{n_fleet}{
    name={\ensuremath{n_{\text{fleet}}}},
    sort={n fleet},
    description={number of vehicles in fleet}
}

\newglossaryentry{road_node}
{
	name={\ensuremath{v}},
	description={road vertex}
}

\newglossaryentry{road_node_post}
{
	name={\ensuremath{w}},
	description={xx}
}

\newglossaryentry{road_node_pre}
{
	name={\ensuremath{u}},
	description={xx}
}

\newglossaryentry{road_edge}
{
	name={\ensuremath{\gls{road_node},\gls{road_node_post}}},
	description={road arc}
}

\newglossaryentry{d}
{
	name={\ensuremath{d}},
	description={distance}
}

\newglossaryentry{d_edge}
{
	type=amod,
	name={\ensuremath{\gls{d}_{\gls{road_edge}}}},
	sort={d v w},
	description={distance of road arc $(\gls{road_edge}) \in \gls{road_edge_set}$}
}

\newglossaryentry{t_edge}
{
	type=amod,
	name={\ensuremath{\gls{t}_{\gls{road_edge}}}},
	sort={t v w},
	description={time to traverse road arc $(\gls{road_edge}) \in \gls{road_edge_set}$}
}

\newglossaryentry{c_unit}
{
	type=amod,
	name={\ensuremath{E_{\gls{c}}}},
	sort={E c},
	description={amount of energy in a charge level}
}

\newglossaryentry{c_edge}
{
	type=amod,
	name={\ensuremath{\gls{c}_{\gls{road_edge}}}},
	sort={c v w},
	description={energy consumption for traversing road arc $(\gls{road_edge})\in\gls{road_edge_set}$}
}

\newglossaryentry{road_graph}
{
	name={\ensuremath{R}},
	description={road graph}
}

\newglossaryentry{road_node_set}
{
	type=amod,
	name={\ensuremath{\mathcal{V}_{\gls{road_graph}}}},
	sort={V R},
	description={set of road vertices}
}

\newglossaryentry{road_edge_set}
{
	type=amod,
	sort={A R},
	name={\ensuremath{\mathcal{A}_{\gls{road_graph}}}},
	description={set of road arcs}
}

\newglossaryentry{exp_network}
{
	name={\ensuremath{G}},
	description={expanded network}
}

\newglossaryentry{exp_node}
{
	name={\ensuremath{\bm{\mathrm{\gls{road_node}}}}},
	description={expanded network node}
}

\newglossaryentry{exp_node_set}
{
	type=amod,
	name={\ensuremath{\mathcal{V}}},
	sort={V},
	description={set of expanded graph vertices}
}

\newglossaryentry{exp_node_post}
{
	name={\ensuremath{\bm{\mathrm{\gls{road_node_post}}}}},
	description={xx}
}

\newglossaryentry{exp_node_pre}
{
	name={\ensuremath{\bm{\mathrm{\gls{road_node_pre}}}}},
	description={xx}
}

\newglossaryentry{exp_edge}
{
	name={\ensuremath{\gls{exp_node},\gls{exp_node_post}}},
	description={expanded network arc}
}

\newglossaryentry{exp_edge_prime}
{
	name={\ensuremath{\gls{exp_node}',\gls{exp_node_post}'}},
	description={expanded network arc}
}

\newglossaryentry{exp_edge_pre}
{
	name={\ensuremath{\gls{exp_node_pre},\gls{exp_node}}},
	description={expanded network arc}
}

\newglossaryentry{exp_node_road}
{
	type=amod,
	name={\ensuremath{\gls{road_node}_{\gls{exp_node}}}},
	sort={v v},
	description={road vertex associated to expanded vertex $\gls{exp_node} \in \gls{exp_node_set}$}
}

\newglossaryentry{exp_edge_road}
{
	name={\ensuremath{\gls{exp_node_road},\gls{exp_node_road_post}}},
	description={xx}
}

\newglossaryentry{exp_node_t}
{
	type=amod,
	name={\ensuremath{\gls{t}_{\gls{exp_node}}}},
	sort={t v},
	description={time step associated to expanded vertex $\gls{exp_node} \in \gls{exp_node_set}$}
}

\newglossaryentry{exp_node_c}
{
	type=amod,
	name={\ensuremath{\gls{c}_{\gls{exp_node}}}},
	sort={c v},
	description={\gls{soc} associated to expanded vertex $\gls{exp_node} \in \gls{exp_node_set}$}
}

\newglossaryentry{exp_node_road_post}
{
	name={\ensuremath{\gls{road_node}_{\gls{exp_node_post}}}},
	description={xx}
}

\newglossaryentry{exp_node_t_post}
{
	name={\ensuremath{\gls{t}_{\gls{exp_node_post}}}},
	description={xx}
}

\newglossaryentry{exp_node_c_post}
{
	name={\ensuremath{\gls{c}_{\gls{exp_node_post}}}},
	description={xx}
}

\newglossaryentry{exp_edge_set}
{
	type=amod,
	name={\ensuremath{\mathcal{A}}},
	sort={A},
	description={set of expanded graph arcs}
}

\newglossaryentry{exp_edge_set_road}
{
	type=amod,
	name={\ensuremath{\gls{exp_edge_set}_{T}}},
	sort={A T},
	description={\glsdesc{exp_edge_set} representing a physical time-dependent movement in the road network}
}

\newglossaryentry{exp_edge_set_charger}
{
	type=amod,
	name={\ensuremath{\gls{exp_edge_set}_{S}}},
	sort={A S},
	description={\glsdesc{exp_edge_set} representing a recharging process}
}

\newglossaryentry{exp_edge_set_idle}
{
	type=amod,
	name={\ensuremath{\gls{exp_edge_set}_{I}}},
	sort={A I},
	description={\glsdesc{exp_edge_set} representing vehicle idling}
}

\newglossaryentry{charger}
{
	name={\ensuremath{s}},
	description={charging station}
}

\newglossaryentry{n_charger}
{
	name={\ensuremath{|\gls{charger_set}|}},
	description={xx}
}

\newglossaryentry{charger_set}
{
	type=amod,
	name={\ensuremath{\mathcal{S}}},
	sort={S},
	description={set of chargers in the road network}
}

\newglossaryentry{charger_cap}
{
	type=amod,
	name={\ensuremath{\bar{S}_{\gls{charger}}}},
	sort={S c},
	description={number of charging plugs at \glsdesc{charger} $\gls{charger} \in \gls{charger_set}$}
}

\newglossaryentry{charger_cap2}
{
	name={\ensuremath{\bar{S}}},
	sort={S},
	description={xx}
}

\newglossaryentry{charger_road}
{
	type=amod,
	name={\ensuremath{\gls{road_node}_{\gls{charger}}}},
	sort={v s},
	description={road vertex associated to station $\gls{charger} \in \gls{charger_set}$}
}

\newglossaryentry{f}
{
	name={\ensuremath{f}},
	description={network flow}
}

\newglossaryentry{f_0}
{
	type=amod,
	name={\ensuremath{\gls{f}_0}},
	sort={f 0},
	description={\glsdesc{f} for rebalancing vehicles}
}

\newglossaryentry{f_m}
{
	type=amod,
	name={\ensuremath{\gls{f}_{\gls{m}}}},
	sort={f m},
	description={\glsdesc{f} for customer trip request $\gls{m} \in \gls{m_set}$}
}

\newglossaryentry{m}
{
	name={\ensuremath{m}},
	description={customer trip request}
}

\newglossaryentry{n_m}
{
	name={\ensuremath{M}},
	sort={M},
	description={number of customer trip requests}
}

\newglossaryentry{m_set}
{
	type=amod,
	name={\ensuremath{\mathcal{M}}},
	sort={M},
	description={set of customer trip requests}
}

\newglossaryentry{dem_m}
{
	type=amod,
	name={\ensuremath{\lambda_{\gls{m}}}},
	sort={zz l m},
	description={customer demand volume of trip request $\gls{in_m} \in \gls{m_set}$}
}

\newglossaryentry{p}{
    name={\ensuremath{p}},
    sort={p},
    description={price}
}

\newglossaryentry{p_dist}{
    type=amod,
    name={\ensuremath{\gls{p}_\text{dist}}},
    sort={p dist},
    description={per-distance driving price}
}

\newglossaryentry{p_energy}{
    type=amod,
    name={\ensuremath{\gls{p}_\text{energy}}},
    sort={p energy},
    description={electricity energy price}
}

\newglossaryentry{p_demand}{
    type=amod,
    name={\ensuremath{\gls{p}_\text{demand}}},
    sort={p demand},
    description={electricity demand charge}
}

\newglossaryentry{p_fleet}{
    type=amod,
    name={\ensuremath{\gls{p}_\text{fleet}}},
    sort={p fleet},
    description={fleet procurement price}
}

\newglossaryentry{p_stn}{
    type=amod,
    name={\ensuremath{\gls{p}_\text{stn}}},
    sort={p infra},
    description={charging station procurement price}
}

\newglossaryentry{rate}{
    name={\ensuremath{\delta}},
    sort={zz d},
    description={charging rate}
}

\newglossaryentry{charger_rate}{
    type=amod,
    name={\ensuremath{\gls{rate}_{\gls{charger}}}},
    sort={zz d s},
    description={charging rate of station $\gls{charger}\in\gls{charger_set}$}
}

\newglossaryentry{rate_set}{
    type=amod,
    name={\ensuremath{\mathcal{D}_{\gls{road_node}}}},
    sort={zz D v},
    description={ordered set (least to greatest) of charging station rates at location $\gls{road_node}\in\gls{road_node_set}$}
}

\newglossaryentry{pmax}{
    type=amod,
    name={\ensuremath{P^{\gls{max}}_{\gls{road_node}}}},
    sort={P s v},
    description={maximum charging demand at location $\gls{road_node}\in\gls{road_node_set}$}
}

\newglossaryentry{incidence_matrix}{
    name={\ensuremath{A}},
    description={incidence matrix of expanded graph, with 1 as in and -1 as out, nodes by edges}
}

\newglossaryentry{incidence_matrix_out}{
    name={\gls{incidence_matrix}^{\gls{out}}},
    description={incidence matrix of expanded graph with only out flows}
}

\newglossaryentry{incidence_matrix_in}{
    name={\gls{incidence_matrix}^{\gls{in}}},
    description={incidence matrix of expanded graph with only in flows}
}
\newglossaryentry{max}
{
	name={\ensuremath{\mathrm{max}}},
	description={xx}
}

\newglossaryentry{out}
{
	name={\ensuremath{\mathrm{out}}},
	description={xx}
}

\newglossaryentry{in}
{
	name={\ensuremath{\mathrm{in}}},
	description={xx}
}

\def\BibTeX{{\rm B\kern-.05em{\sc i\kern-.025em b}\kern-.08em
    T\kern-.1667em\lower.7ex\hbox{E}\kern-.125emX}}

\title{\LARGE \bf
Joint Optimization of Autonomous Electric Vehicle Fleet Operations and Charging Station Siting
}

\author{Justin Luke$^{1}$, Mauro Salazar$^{2}$, Ram Rajagopal$^{3}$, and Marco Pavone$^{4}$
\thanks{$^{1}$Justin Luke is with the Department of Civil and Environmental Engineering, Stanford University, Stanford, CA 94035, USA
        {\tt\small jthluke@stanford.edu}}%
\thanks{$^{2}$Mauro Salazar is with the Control Systems Technology section, Eindhoven
University of Technology (TU/e), Eindhoven, MB 5600, The Netherlands
        {\tt\small m.r.u.salazar@tue.nl}}%
\thanks{$^{3}$Ram Rajagopal is with the Department of Civil and Environmental Engineering, Stanford University, Stanford, CA 94035, USA
        {\tt\small ramr@stanford.edu}}%
\thanks{$^{4}$Marco Pavone is with the Department of Aeronautics and Astronautics, Stanford University, Stanford, CA 94035, USA
        {\tt\small pavone@stanford.edu}}%
}

\begin{document}

\maketitle
\thispagestyle{empty}
\pagestyle{empty}

\begin{abstract}
Charging infrastructure is the coupling link between power and transportation networks, thus determining charging station siting is necessary for planning of power and transportation systems.
While previous works have either optimized for charging station siting given historic travel behavior, or optimized fleet routing and charging given an assumed placement of the stations, this paper introduces a linear program that optimizes for station siting and macroscopic fleet operations in a joint fashion.
Given an electricity retail rate and a set of travel demand requests, the optimization minimizes total cost for an autonomous EV fleet comprising of travel costs, station procurement costs, fleet procurement costs, and electricity costs, including demand charges.
Specifically, the optimization returns the number of charging plugs for each charging rate (e.g., Level 2, DC fast charging) at each candidate location, as well as the optimal routing and charging of the fleet.
From a case-study of an electric vehicle fleet operating in San Francisco, our results show that, albeit with range limitations, small EVs with low procurement costs and high energy efficiencies are the most cost-effective in terms of total ownership costs.
Furthermore, the optimal siting of charging stations is more spatially distributed than the current siting of stations, consisting mainly of high-power Level 2 AC stations (16.8 kW) with a small share of DC fast charging stations and no standard 7.7kW Level 2 stations.
Optimal siting reduces the total costs, empty vehicle travel, and peak charging load by up to 10\%.
\end{abstract}

\section{Introduction}
\glsresetall
Electrification and vehicle autonomy are driving down the total cost of ownership for vehicle fleets.
Presently, autonomous \glspl{ev} are being developed for fleet applications such as passenger mobility-on-demand services.
The development of these \gls{eamod} fleets are motivated by the low maintenance and energy costs of \glspl{ev}~\cite{palmer2018total}, low operating costs of shared autonomous vehicles~\cite{SpieserTreleavenEtAl2014}, and policy mandates for the decarbonization of the transportation sector~\cite{styczynski2019public}.
Autonomous fleets also have the advantage of highly controllable routing and charge scheduling compared to privately owned human-operated \glspl{ev}.
However, in all these developments, the optimal planning for the charging infrastructure needed to support future mobility fleets at scale is not well understood, considering the intersection of emerging trends in mobility-on-demand services, vehicle electrification, and driving automation.
Crucially, the operation of future \gls{eamod} systems will be strongly influenced by the available charging infrastructure, which in turn should be designed to accommodate the \glspl{ev}' charging activities in the best possible way.
These problems are intimately coupled, calling for optimization methods to systematically solve them.
Against this background, this paper proposes a convex optimization framework to jointly optimize the design of the charging infrastructure and the operation of a centrally controlled \gls{eamod} fleet.

\paragraph{Literature review}
This paper contributes to two main research streams.
The first stream focuses on the routing and charge scheduling problem of fleet \glspl{ev}. Network flow models have been successfully used to minimize fleet travel and electricity costs subject to fulfilment of customer trip requests, limited driving range, and charging constraints imposed by congestion on the power transmission grid~\cite{RossiIglesiasEtAl2018b}, also accounting for the distribution grid~\cite{EstandiaSchifferEtAl2019}.
A vehicle coordination and charge scheduling algorithm is proposed in~\cite{BoewingSchifferEtAl2020} to efficiently optimize the operation of an \gls{eamod} fleet accounting for the battery level of individual vehicles and the
energy availability in the power grid.
A heuristic algorithm for the electric traveling salesman with time windows is developed in~\cite{104ffb30ecec45c397a09d2687d4d6f9} to solve customer routing and recharging in small-scale problems.
However, the optimized operations are determined for an assumed siting layout of charging stations.
The second stream focuses on the design of the charging infrastructure considering the \glspl{ev}' operations as exogenous data. Previous works have largely framed the station siting problem as solving variants of mixed integer linear programs \cite{chen2013electric,6183283,7581114}.
Thereby, the objective terms commonly include user access costs and station construction costs under assumptions of desired user charging behavior, determined from historic origin-destination travel data or parking dwell times.
However, this class of models do not consider the greater charging flexibility of autonomous fleet vehicles, due to their capability to operate after user drop-off and reposition to alternate charging locations.
Furthermore, these combinatorial optimization approaches suffer from computational scalability as their complexity rises significantly with the problem size.
In conclusion, while the former research stream focuses on optimizing the \gls{eamod} activities for a given charging infrastructure, the latter stream bases the infrastructure design problem on historic travel data and frames it as a mixed-integer problem.

\paragraph{Statement of contribution}
This paper bridges the gap between the aforementioned research streams and, rather than considering separate optimization problems, efficiently solves for the station planning and macroscopic fleet operations jointly.
Specifically, we first propose a network flow model describing the \glspl{ev}' routing and charging activities and combine it with the design of the fleet size and the infrastructure siting. Second, we frame the optimal design and control problem minimizing the total cost incurred by the \gls{eamod} operator (defined as the sum of the fleet's routing and charging costs, and the procurement costs for the fleet and infrastructure design) as a linear program that can be efficiently solved with off-the-shelf optimization algorithms.
Finally, we showcase our framework on a case study for San Francisco, CA, where we investigate the impact of the \gls{ev} type on the resulting optimal design and operation, and highlight the importance of jointly optimizing the \gls{eamod} system's routing and charging with the infrastructure siting.

\paragraph{Organization}
The remainder of the paper is structured as follows:
\cref{sec:model} introduces the \gls{eamod} charging station siting joint optimization model.
\cref{sec:case_study} details our case study of San Francisco, CA, and presents results on how the joint optimization varies with \gls{ev} models of differing battery sizes, and how the joint optimization performs when compared to a baseline based upon present-day charging station siting.
We conclude the paper in \cref{sec:conclusion} with a summary of our key findings and an outlook for future research.
\section{Modeling the Joint Optimization of \gls{eamod} Systems and Charging Station Siting}
\label{sec:model}
An \gls{eamod} system consists of electric autonomous fleet vehicles that serve customer travel requests.
When vehicles are not serving customers, they may be recharging or performing rebalancing trips.
Recharging can vary spatially, temporally, and also by charging rate.
Rebalancing trips, defined as vehicle travel without carrying a customer, serve to reposition vehicles to charging stations in advance of charging events, or to resolve spatial and temporal mismatches between the origins and destinations of customer travel requests.

In previous works, the \gls{eamod} system is modeled as a network flow problem, utilizing an expanded road graph that has vertices representing coordinates in location, time, and battery charge level \cite{RossiIglesiasEtAl2018b}\cite{EstandiaSchifferEtAl2019}.
In this section, we will present a new \gls{eamod} model that extends from the original model by introducing electricity demand charges, charge throttling, and the joint optimization of charging station siting.
We will sequentially detail each component of model before presenting the complete \gls{eamod} problem at the end of the section.

\paragraph{Expanded graph representation} We model a transportation network as a directed graph $G_{\gls{road_graph}}=\parentheses{\gls{road_node_set}, \gls{road_edge_set}}$ with a set of vertices $\gls{road_node} \in \gls{road_node_set}$ representing locations and a set of arcs $(\gls{road_edge})\in\gls{road_edge_set}$ representing the route between $\gls{road_node}$ and $\gls{road_node_post}$.
Each arc $(\gls{road_edge})\in\gls{road_edge_set}$ is characterized by the route's distance $\gls{d_edge}$, travel time $\gls{t_edge}$, and energy required to traverse it $\gls{c_edge}$.
In contrast with \cite{RossiIglesiasEtAl2018b} and \cite{EstandiaSchifferEtAl2019}, we allow for self-loop arcs to capture travel demand that begins and ends in the same location, as the locations can represent city regions.

We define $\gls{t_set}=\{1,...,\gls{n_t}\}$ as the set of equidistant time steps of duration $\gls{t_step}\in\reals^+$, and $\gls{c_set}=\{1,...,\gls{n_c}\}$ as the set of equidistant battery charge level discretizations, each with energy $\gls{c_unit}\in\reals^+$. The battery charge levels indicate the \gls{soc} levels of an \gls{ev}, with $c=1$ representing an empty battery and $c=\gls{n_c}$ representing a full battery.

The locations in $\gls{road_node_set}$ represent destinations for customer travel and also points of access to charging stations.
We define $\gls{charger_set}$ as the set of charging stations in the network, with each station $\gls{charger}\in\gls{charger_set}$ defined by a tuple $\gls{charger}=\parentheses{\gls{charger_road}, \gls{charger_rate}}$ where $\gls{charger_road}\in\gls{road_node_set}$ is the station's location and $\gls{charger_rate}\in\mathbb{R}^+$ is the station's per-plug charging rate.
Additionally, each station $\gls{charger}\in\gls{charger_set}$ has a number of plugs, $\gls{charger_cap}\in\mathbb{R}^+$.
Note that in contrast with \cite{RossiIglesiasEtAl2018b} and \cite{EstandiaSchifferEtAl2019}, this model allows for multiple stations with different charging rates to be situated at the same location.

We then define the directed multigraph $\gls{exp_network}=(\gls{exp_node_set}, \gls{exp_edge_set})$ as the expanded transportation network, which expands $G_{\gls{road_graph}}$ along the dimensions of time and battery charge level.
The vertex set $\gls{exp_node_set}\subseteq \gls{road_node_set} \times \gls{t_set} \times \gls{c_set}$ contains vertices $\gls{exp_node}\in\gls{exp_node_set}$ that are defined by the tuple $(\gls{exp_node_road}, \gls{exp_node_t}, \gls{exp_node_c})$ in which entries specify location, time, and \gls{soc}, respectively.

The arc set $\gls{exp_edge_set}$ is the union of two disjoint subsets, $\gls{exp_edge_set_road}\cup\gls{exp_edge_set_charger}$. Travel in the transportation network is represented by arcs $(\gls{exp_edge})\in\gls{exp_edge_set_road}$ and is defined by
\begin{align*}
    \gls{exp_edge_set_road} = &\{(\gls{exp_edge})\in\gls{exp_edge_set} \mid (\gls{exp_edge_road})\in\gls{road_edge_set}, \\
    & \gls{exp_node_t_post} - \gls{exp_node_t} = \gls{t}_{\gls{exp_edge_road}}, \gls{exp_node_c} - \gls{exp_node_c_post} = \gls{c}_{\gls{exp_edge_road}} \}.
\end{align*}
This definition enforces that the time expansion for travel from $\gls{exp_node}$ to $\gls{exp_node_post}$ is equal to the travel time $\gls{t}_{\gls{exp_edge_road}}$ and charge expansion is equal to the travel charge $\gls{c}_{\gls{exp_edge_road}}$. 
Unlike in $G_{\gls{road_graph}}$, the distance, travel time, and travel energy between two locations in $\gls{exp_network}$ can be time-varying.
Idling vehicles, which only move forward in time from $\gls{exp_node}$ to $\gls{exp_node_post}$ but remain fixed in location and \gls{soc}, are a subset of $\gls{exp_edge_set_road}$:
\begin{equation*}
    \gls{exp_edge_set_idle} =
    \{(\gls{exp_edge})\in\gls{exp_edge_set_road} \mid \gls{exp_node_road}=\gls{exp_node_road_post},\gls{exp_node_t_post}=\gls{exp_node_t} + 1, \gls{exp_node_c}=\gls{exp_node_c_post}\}
\end{equation*}

Recharging at charging stations is represented by arcs $(\gls{exp_edge})\in\gls{exp_edge_set_charger}$ and is defined by
\begin{align*}
    \gls{exp_edge_set_charger} = &\{(\gls{exp_edge})\in\gls{exp_edge_set} \mid \gls{exp_node_road}=\gls{exp_node_road_post}=\gls{charger_road}\quad \forall\gls{charger}\in\gls{charger_set},\\
    &\gls{exp_node_t_post} - \gls{exp_node_t} = 1,\\
    &\frac{(\gls{exp_node_c_post} - \gls{exp_node_c})\gls{c_unit}}{\gls{t_step}} = \gls{rate}, \\
    &\gls{rate}\in\{\frac{\gls{c_unit}}{\gls{t_step}},2\frac{\gls{c_unit}}{\gls{t_step}},...,\gls{charger_rate}\} \}.
\end{align*}
This definition enforces that for the recharging process represented by the arc $(\gls{exp_edge})$, the location is fixed to the location of a station $\gls{charger}$, the recharging occurs over one time step, and the charge rate derived from the arc's charge expansion is at most the charge rating of the station.
In contrast with \cite{RossiIglesiasEtAl2018b} and \cite{EstandiaSchifferEtAl2019}, we allow for charging at station $\gls{charger}$ to include rates below its rated capacity in order to model charge throttling, a feature that is becoming increasingly common in modern charging stations and \glspl{ev} that allows for greater control over power demand \cite{tan2016integration}.
Note that at locations with multiple charging stations, multiple edges will be defined for a given pair $(\gls{exp_edge})$, thereby making $\gls{exp_network}$ a multigraph.

\paragraph{Customer travel requests} We define $\gls{m_set} = \{1, ..., \gls{n_m}\}$ as the set of travel requests which the \gls{eamod} fleet must serve.
Each request $\gls{m}\in\gls{m_set}$ is defined by a tuple $\gls{m} = (\gls{road_node}_{\gls{m}},\gls{road_node_post}_{\gls{m}},\gls{t}_{\gls{m}},\gls{dem_m})\in\gls{road_node_set}\times\gls{road_node_set}\times\gls{t_set}\times\reals^+$ in which the entries represent the origin, destination, departure time, and travel demand volume, respectively.
We define
\begin{equation*}
    \gls{indicator}(\gls{exp_edge}, \gls{m}) = \gls{indicator}_{(\gls{exp_node_road}, \gls{exp_node_road_post}, \gls{exp_node_t})=(\gls{road_node}_{\gls{m}}, \gls{road_node_post}_{\gls{m}}, \gls{t}_{\gls{m}})}.
\end{equation*}
as an indicator for whether arc $(\gls{exp_edge})\in\gls{exp_edge_set}$ fulfils request $\gls{m}$.
The distance, travel time, and travel energy of each arc $(\gls{exp_edge})\in\gls{exp_edge_set_road}$ is consistent with the distance, travel time, and travel energy of the travel request $\gls{m}$ which the arc fulfills.

\paragraph{Vehicle flows} In the \gls{eamod} network flow problem, we solve for vehicle flows on the expanded graph \gls{exp_network}.
We define $\gls{f}(\gls{exp_edge}) : \gls{exp_edge_set}\rightarrow\reals^+$ to represent vehicle flow on arc $(\gls{exp_edge})$.
The activity of the vehicles depends upon the arc $(\gls{exp_edge})$.
If $(\gls{exp_edge})\in\gls{exp_edge_set_idle}$, then vehicles are idling.
If $(\gls{exp_edge})\in\gls{exp_edge_set_charger}$, then the vehicles are recharging.
If $(\gls{exp_edge})\in\gls{exp_edge_set_road} - \gls{exp_edge_set_idle}$, then vehicles could be carrying customers or rebalancing.
In the \gls{eamod} problem presented in this paper, the fleet must serve all customer travel requests as a hard constraint.
Therefore, any travel along $(\gls{exp_edge})\in\gls{exp_edge_set_road}$ in excess of the customer travel demand volume must be rebalancing flow.
We define $\gls{f_0}(\gls{exp_edge}) : \gls{exp_edge_set}\rightarrow\reals^+$ to represent vehicle rebalance flow on arc $(\gls{exp_edge})$.
Note that $\gls{f_0}(\gls{exp_edge}) \leq \gls{f}(\gls{exp_edge}) \quad \forall (\gls{exp_edge})\in\gls{exp_edge_set}$ as rebalancing flow on a given arc must be a fraction of the total flow of that arc.
Once \gls{f} is determined, the rebalancing flow $\gls{f_0}$ can be recovered post-optimization. For every $(\gls{exp_edge})\in\gls{exp_edge_set_road}$ and $\gls{m}\in\gls{m_set}$, if $\gls{indicator}(\gls{exp_edge}, \gls{m})=1$, we compute
\begin{align*}
    \gls{f_0}(\gls{exp_edge}) &= \max(\csum{(\gls{exp_edge_prime})\in\gls{exp_edge_set_road} - \gls{exp_edge_set_idle}:\gls{indicator}(\gls{exp_edge_prime}, \gls{m})=1}{\gls{f}(\gls{exp_edge_prime})} \\
    &- \gls{dem_m}, 0)\parentheses{\csum{(\gls{exp_edge_prime})\in\gls{exp_edge_set_road} - \gls{exp_edge_set_idle}:\gls{indicator}(\gls{exp_edge_prime}, \gls{m})=1}{\gls{f}(\gls{exp_edge_prime})}}^{-1}\gls{f}(\gls{exp_edge}).
\end{align*}
Here, we assume rebalancing travel is distributed uniformly among the arcs satisfying $\gls{indicator}(\gls{exp_edge}, \gls{m})=1$.
For any arcs $(\gls{exp_edge})\in\gls{exp_edge_set_road}$ that do not correspond to any travel request, $\gls{f_0}(\gls{exp_edge}) = \gls{f}(\gls{exp_edge})$.  

\paragraph{Electricity demand charges} To introduce the modeling of electricity demand charges incurred by the fleet for their peak charging load at each location, we define \gls{pmax}:
\begin{equation}
    \label{eq:pmax}
    \gls{pmax} = \max_{\gls{t}\in\gls{t_set}}{\csum{(\gls{exp_edge})\in\gls{exp_edge_set_charger}}{\gls{indicator}_{(\gls{exp_node_road}, \gls{exp_node_t})=(\gls{road_node}, \gls{t})}\gls{f}(\gls{exp_edge})\frac{(\gls{exp_node_c_post} - \gls{exp_node_c})\gls{c_unit}}{\gls{t_step}}}} \quad\forall \gls{road_node}\in\gls{road_node_set}
\end{equation}
Here, we compute the total charging demand at location $\gls{road_node}$ and time $\gls{t}$ by including all associated charging arcs and taking the sum product of vehicle flow of the arc and the arc's charge expansion, with appropriate conversions for units of power. We then take the maximum over all times to determine $\gls{pmax}$.

\paragraph{Charging station siting} To introduce optimal sizing of stations of varying charging rate and location, we make the number of plugs $\gls{charger_cap}$ at each station $\gls{charger}\in\gls{charger_set}$ an optimization variable in the \gls{eamod} problem.
Because our model allows for charge throttling, the determination of number of plugs for slower charging rate stations is dependent upon the residual capacity of any co-located faster rate stations.
To assist with the derivation of the equations that determine charging station sizing, we use an example.

First, we define $\gls{rate_set}$ as the \glsdesc{rate_set}.
Let $\gls{rate_set}(j)$ return the $j^{\text{th}}$ element of $\gls{rate_set}$ for all $j\in\{1,...,\myabs{\gls{rate_set}}\}$, with $\gls{rate_set}(0) = \unit[0]{kW}$.
We will also define the following set, which includes all charging arcs representing charging at location $\gls{road_node}$ at time $\gls{t}$ charging at a rate that is greater than $\gls{rate}_{1}$ but at most $\gls{rate}_{2}$:
\begin{align*}
    \gls{exp_edge_set_charger}^{\gls{road_node}, \gls{t}, \gls{rate}_{1}, \gls{rate}_{2}} = &\{(\gls{exp_edge})\in\gls{exp_edge_set_charger} \mid \gls{exp_node_road} = \gls{road_node},\\ &\gls{exp_node_t} = \gls{t}, \gls{rate}_{1} < \frac{(\gls{exp_node_c_post} - \gls{exp_node_c})\gls{c_unit}}{\gls{t_step}} \leq \gls{rate}_{2}\}
\end{align*}
In this example, consider a location $\gls{road_node}$ with the option to install charging stations of three charging rates: $\gls{rate_set} = \{\unit[7.7]{kW}, \unit[16.8]{kW}, \unit[50.0]{kW}\}$.
Only the $\unit[50.0]{kW}$ station can charge at the rates from $(\unit[16.8]{kW},\unit[50.0]{kW}]$, thus the constraint on charging flows with charging rates in this range is:
\begin{equation*}
    \csum{(\gls{exp_edge})\in \gls{exp_edge_set_charger}^{\gls{road_node}, \gls{t}, 16.8, 50.0}}{\gls{f}(\gls{exp_edge})} \leq \gls{charger_cap2}_{(\gls{road_node}, 50.0)}
\end{equation*}
Following, the constraint on charging flows with rates in $(\unit[7.7]{kW},\unit[16.8]{kW}]$ is given by the number of plugs of the $\unit[16.8]{kW}$ station plus any unused plugs at the $\unit[50.0]{kW}$ station:
\begin{equation*}
    \csum{(\gls{exp_edge})\in \gls{exp_edge_set_charger}^{\gls{road_node}, \gls{t}, 7.7, 16.8}}{\gls{f}(\gls{exp_edge})} \leq \gls{charger_cap2}_{(\gls{road_node}, 16.8)}
    + (\gls{charger_cap2}_{(\gls{road_node}, 50.0)} - \csum{(\gls{exp_edge})\in \gls{exp_edge_set_charger}^{\gls{road_node}, \gls{t}, 16.8, 50.0}}{\gls{f}(\gls{exp_edge})})
\end{equation*}
Similarly, the constraint for flows with rates in $(\unit[0]{kW},\unit[7.7]{kW}]$ is given by:
\begin{align*}
    \csum{(\gls{exp_edge})\in \gls{exp_edge_set_charger}^{\gls{road_node}, \gls{t}, 0, 7.7}}{\gls{f}(\gls{exp_edge})} \leq \gls{charger_cap2}_{(\gls{road_node}, 7.7)}
    &+ (\gls{charger_cap2}_{(\gls{road_node}, 50.0)} - \csum{(\gls{exp_edge})\in \gls{exp_edge_set_charger}^{\gls{road_node}, \gls{t}, 16.8, 50.0}}{\gls{f}(\gls{exp_edge})})\\
    &+ (\gls{charger_cap2}_{(\gls{road_node}, 16.8)} - \csum{(\gls{exp_edge})\in \gls{exp_edge_set_charger}^{\gls{road_node}, \gls{t}, 7.7, 16.8}}{\gls{f}(\gls{exp_edge})})
\end{align*}
After collecting like terms, the general form of the equations that determine charging station sizing are:
\begin{align}
    \label{eq:infra_constr}
    &\sum_{i=j}^{\myabs{\gls{rate_set}}}(\csum{\substack{(\gls{exp_edge})\in\\ \gls{exp_edge_set_charger}^{\gls{road_node}, \gls{t}, \gls{rate_set}(i-1), \gls{rate_set}(i)}}}{\gls{f}(\gls{exp_edge})} - \gls{charger_cap2}_{(\gls{road_node}, \gls{rate_set}(i))}) \leq 0 \\
    &\forall j\in\{1,...,\myabs{\gls{rate_set}}\}, \forall\gls{road_node}\in\gls{road_node_set}, \forall\gls{t}\in\gls{t_set}\nonumber
\end{align}

\paragraph{\gls{eamod} model with station siting} The cost terms that are considered in the \gls{eamod} problem are fleet procurement, charging station procurement, electricity (both energy consumption and demand charges), and vehicle maintenance.
We seek to jointly solve for the vehicle flows $\gls{f}(\gls{exp_edge})\quad\forall(\gls{exp_edge})\in\gls{exp_edge_set}$ and charging station sizing $\gls{charger_cap}\quad\forall\gls{charger}=(\gls{road_node},\gls{rate})\in\gls{charger_set},\forall \gls{road_node}\in\gls{road_node_set},\forall\gls{rate}\in\gls{rate_set}$ such that total fleet costs are minimized:

\begin{mini!}<b>
	{\substack{\gls{f}, [\gls{charger_cap}]_{\gls{charger}\in\gls{charger_set}}, \\
			[\gls{pmax}]_{\gls{road_node}\in\gls{road_node_set}}, \gls{n_fleet}}}
	{\csum{(\gls{exp_edge})\in\gls{exp_edge_set_charger}}{\gls{f}(\gls{exp_edge})(\gls{exp_node_c_post} - \gls{exp_node_c})\gls{c_unit}\gls{p_energy}(\gls{exp_node_t})}}
	{\label{eq:joint_eamod_prob}}{}
	\breakObjective {+ 
		\csum{\gls{road_node}\in\gls{road_node_set}}\gls{pmax}\gls{p_demand}}
	\breakObjective {+ 
		\csum{(\gls{exp_edge})\in\gls{exp_edge_set_road}-\gls{exp_edge_set_idle}}{\gls{f}(\gls{exp_edge})\gls{d}_{\gls{exp_edge_road}}\gls{p_dist}}}
	\breakObjective{+ \gls{n_fleet}\gls{p_fleet}}
	\breakObjective {+ 
		\csum{\gls{road_node}\in\gls{road_node_set}}\csum{j=1}^{\myabs{\gls{rate_set}}}\gls{charger_cap2}_{(\gls{road_node}, \gls{rate_set}(j))}\gls{p_stn}(\gls{rate_set}(j)) \label{eq:eamod_infra_problem_obj}}
	%
	\addConstraint{
		\mathbb{1}\T (-\gls{incidence_matrix_out}[\gls{exp_node_set}_1, :]\gls{f}) = \gls{n_fleet}
		\label{eq:fleet_size}}	
	\addConstraint{
		\csum{\substack{(\gls{exp_edge})\\\in\gls{exp_edge_set_road} - \gls{exp_edge_set_idle}}}{\gls{f}(\gls{exp_edge})\gls{indicator}(\gls{exp_edge}, \gls{m})} = \gls{dem_m} \quad\forall \gls{m}\in\gls{m_set} \label{eq:trip_flow}}
	%
	\addConstraint{
		\gls{incidence_matrix_in} [\gls{exp_node_set}_{2:\gls{n_t}-1}, :]\gls{f} = -\gls{incidence_matrix_out} [\gls{exp_node_set}_{2:\gls{n_t}-1}, :]\gls{f}	 \label{eq:fleet_dyn_1}}
	\addConstraint{
		\gls{incidence_matrix_in} [\gls{exp_node_set}_{\gls{n_t}}, :]\gls{f} = -\gls{incidence_matrix_out} [\gls{exp_node_set}_1, :]\gls{f} \label{eq:fleet_dyn_2}}
	\addConstraint{
		\csum{(\gls{exp_edge})\in\gls{exp_edge_set_charger}}{\gls{indicator}_{(\gls{exp_node_road}, \gls{exp_node_t})=(\gls{road_node}, \gls{t})}\gls{f}(\gls{exp_edge})\frac{(\gls{exp_node_c_post} - \gls{exp_node_c})\gls{c_unit}}{\gls{t_step}}}\nonumber
		}
	\addConstraint{
		\quad\leq \gls{pmax} \quad\forall\gls{road_node}\in\gls{road_node_set}, \forall\gls{t}\in\gls{t_set} \label{eq:pmax_constr}}
	\addConstraint{
        \cref{eq:infra_constr}\label{eq:infra_constr_opt}}
\end{mini!}


Objective term (3a) is the electricity energy cost, with time-varying electricity price $\gls{p_energy}(\gls{t}):\gls{t_set}\rightarrow\reals^+$ having units of $\unitfrac[]{USD}{kWh}$.
Objective term (3b) is the electricity demand cost, with demand charge $\gls{p_demand}\in\reals^+$ having units of $\unitfrac[]{USD}{kW}$.
Maximum power demand at each location $\gls{pmax}$ is defined from $\gls{f}$ in \cref{eq:pmax_constr}, which follows from the derivation of \cref{eq:pmax}.
Objective term (3c) is the maintenance cost due to vehicle travel, with $\gls{p_dist}\in\reals^+$ representing maintenance cost per distance traveled and has units of $\unitfrac[]{USD}{km}$.
Objective term (3d) is the fleet procurement cost, with vehicle price $\gls{p_fleet}\in\reals^+$ having units of $\unitfrac[]{USD}{vehicle}$.
Fleet size $\gls{n_fleet}\in\reals^+$ is defined from $\gls{f}$ in \cref{eq:fleet_size}. Lastly, objective term (3e) is the charging station procurement cost, with $\gls{p_stn}(\gls{rate}) : \gls{rate_set}\rightarrow\reals^+$ having units of $\unitfrac[]{USD}{station}$.

We introduce $\gls{incidence_matrix}\in\reals^{\myabs{\gls{exp_node_set}}}\times\reals^{\myabs{\gls{exp_edge_set}}}$, the incidence matrix of the expanded graph \gls{exp_network}, in which +1 entries indicate into node and -1 indicates out of node.
$\gls{incidence_matrix_out}$ is the incidence matrix with only the -1 entries, whereas $\gls{incidence_matrix_in}$ is the incidence matrix with only the +1 entries.
We denote $\gls{incidence_matrix} [ \mathcal{V}_{\gls{t}}, : ]$ as the selection of all columns and the rows corresponding to nodes in $\mathcal{V}_{\gls{t}}$, in which $\mathcal{V}_{\gls{t}}$ is defined as the set of all expanded graph nodes associated with time $\gls{t}$.
Similarly, $\mathcal{V}_{\gls{t}_{1}:\gls{t}_{2}}$ returns the set of expanded graph nodes associated with a range of times, from $\gls{t}_{1}$ to $\gls{t}_{2}$.

\cref{eq:trip_flow} enforces the fleet to serve all customer travel requests.
\cref{eq:fleet_dyn_1} enforces consistency and flow conservation in the network flow problem, as detailed in \cite{RossiIglesiasEtAl2018b} and \cite{EstandiaSchifferEtAl2019}.
\cref{eq:fleet_dyn_2} enforces a periodicity constraint, ensuring vehicles return to a state in the final time that is identical to its state in the initial time.
\cref{eq:infra_constr_opt} determines the station sizing variables $\gls{charger_cap}$ and enforces station plug capacities, as derived in \cref{eq:infra_constr}.

The \gls{eamod} problem \cref{eq:joint_eamod_prob} is a linear program that is amenable for any general LP solver.
It has $\myabs{\gls{road_edge_set}}\myabs{\gls{c_set}}\myabs{\gls{t_set}} + \myabs{\gls{road_node_set}}\myabs{\gls{rate_set}} + \myabs{\gls{road_node_set}} + 1$ decision variables.
The dominant term is $\myabs{\gls{road_edge_set}}\myabs{\gls{c_set}}\myabs{\gls{t_set}}$.
Given that $\gls{road_edge_set}$ can be at most $\myabs{\gls{road_node_set}}^2$ if there is a route between every location, it follows that $\myabs{\gls{road_edge_set}}\myabs{\gls{c_set}}\myabs{\gls{t_set}}\in\mathcal{O}(\myabs{\gls{road_node_set}}^2\myabs{\gls{c_set}}\myabs{\gls{t_set}})$.

A few comments are in order.
First, the transportation graph $G_{\gls{road_graph}}$ uses routes between destinations as arcs, as opposed to road segments.
The advantage of this modeling choice is that origin-destination travel demand data sets, with data on the traffic volume, distance, and travel time of the route connecting an origin-destination pair, can be readily used to formulate this graph; no assumptions are needed about how a vehicle traverses physical road segments in its routing.
The disadvantage of this approach is that traffic congestion along road segments cannot be accurately modeled.
However, if road topography data is available and shortest-path routing can be assumed, then \cref{eq:joint_eamod_prob} can easily be adapted to use road segments and enforce threshold congestion constraints in the optimization, as presented in \cite{EstandiaSchifferEtAl2019}.
Second, the flow solved by \cref{eq:joint_eamod_prob} allows for fractional flows of vehicles.
However, this is acceptable given the macroscopic nature of the station siting problem, as arc flows are on the order of thousands of vehicles and a fractional vehicle has negligible impact on results.
Third, although the number of decision variables scales with $\mathcal{O}(\myabs{\gls{road_node_set}}^2\myabs{\gls{c_set}}\myabs{\gls{t_set}})$, if the operating region of the fleet remains fixed, increasing the number of locations will increase the spatial granularity which requires a finer charge and time discretization.
In such a case, the scaling would be more than quadratic with the number of locations.
\section{Case Study in San Francisco, CA}
\label{sec:case_study}
\subsection{Model parameters}
\label{sec:case_study_params}
To investigate how charging stations are sited when optimized jointly with \gls{eamod} fleet operations, we conduct a detailed case study of an \gls{eamod} system in San Francisco, CA.
The fleet operates for 24 hours, serving all travel demand in the city on a typical weekday.
The fleet additionally serves San Francisco customers traveling to and from North Bay, East Bay, and the Peninsula.
The fleet operator installs charging stations within San Francisco at which fleet vehicles have exclusive use for recharging; these installations are determined by the joint charging station siting optimization.
To evaluate the impact of the vehicle battery size and efficiency, we conduct separate optimizations for the following three \gls{ev} models: the \glsdesc{model3} ($\unit[75]{kWh}$ battery), the \glsdesc{leaf} ($\unit[40]{kWh}$), and the \glsdesc{spring} ($\unit[27.4]{kWh}$).

\paragraph{Data for expanded graph} We use travel demand data provided by StreetLight Data from 2019, averaged across Mondays-Fridays.
The hourly origin-destination data is given for the 190 \gls{taz} of San Francisco \cite{mtc} and three pass-through zones measuring traffic flow across the Golden Gate Bridge (to/from North Bay), the Bay Bridge (to/from East Bay), and all roads connecting the southern city boundary to the Peninsula.
The 190 \glspl{taz} are then aggregated into 25 city zones determined by dividing San Francisco into a 5x5 grid of 2.2 km x 2.2 km cells.
With 25 city zones and three pass-through zones, we have $\myabs{\gls{road_node_set}} = 28$.
In total, the \gls{eamod} serves 2.89 million customer requests over the weekday.
The distance and travel time between origin-destination pairs are also given by this data set.
To account for the travel that occurs beyond the city boundaries, we add 32 km, 24 km, and 48 km to the distance and 40 min, 20 min, and 45 min to the travel time for vehicles traveling to or from the North Bay, East Bay, and Peninsula pass-through zones, respectively.
To determine the energy needed to travel between an origin-destination pair, we assume the \glspl{ev} consume energy at the rate of their combined cycle energy efficiency rating, as rated by the US \gls{epa} for the \gls{model3} and \gls{leaf} and the \gls{wltp} for the \gls{spring}.
These ratings are provided in \cref{table:evs}.
We can then determine travel energy directly from the travel distance data.

The 24-hour horizon is discretized into fifteen-minute time steps, such that $\myabs{\gls{t_set}} = 96$.
This time discretization was chosen to be close to the travel time of the shortest duration trip requests.
It is also consistent with the frequency at which power is metered by the electric utility, which is how the utility determines energy and demand charges.
We assume the hourly travel demand is uniformly distributed over the hour. For example, if the number of customer requests from 07:00 to 08:00 is 1000 for a particular origin and destination pair, we assume the number of requests at the time steps 07:00, 07:15, 07:30, and 07:45 to be 250.

We conservatively assume the \glspl{ev} operate with their \gls{soc} between 0.2 and 0.8. This simple policy accounts for having emergency reserve battery charge, avoiding battery degradation effects at high \glspl{soc} \cite{lunz2012influence}, and the capacity loss of the battery over its usage.
The \gls{soc}-restricted vehicle battery is then discretized into energy steps of $\unit[0.74]{kWh}$, such that $\myabs{\gls{c_set}} = 62, 33,$ and $23$ for the \gls{model3}, \gls{leaf}, and \gls{spring}, respectively.
This charge discretization was chosen to be close to the travel energy of the lowest energy trip requests.

The fleet operator can install charging stations at any of the 25 city zones.
There are four different station options with varying charging rates: two Level 2 AC options, $\unit[7.7]{kW}$ and $\unit[16.8]{kW}$, and two DC fast charging options, $\unit[50]{kW}$ and $\unit[150]{kW}$.
These options were chosen from common station configurations used in present day \cite{rmi}.
However, as the \gls{spring} has a smaller battery, the maximum charging rate it can receive is $\unit[30]{kW}$.
Thus, for the \gls{spring} optimization, although DC fast charging stations are considered, their output charging rates are limited to $\unit[30]{kW}$.
All \glspl{ev} are assumed to recharge with a grid-to-battery efficiency of 90\%.

\paragraph{Pricing data} We set the per-distance maintenance cost of \gls{ev} travel to $\gls{p_dist} = \unitfrac[0.0464]{USD}{km}$ ($\unitfrac[0.0746]{USD}{mi}$) in accordance with the \gls{aaa}\cite{aaa}.

For electricity pricing, we use the ``Business Electric Vehicle'' electric schedule from \gls{pge}, the primary electric utility that serves San Francisco \cite{pge}.
This schedule consists of a \gls{tou} rate, $\gls{p_energy}\parentheses{\gls{t}}$, a price per unit energy that varies by time of day, and a subscription-based demand charge, $\unitfrac[85.98]{USD}{50 kW}$, a price per unit power of maximum load within a billing month.
The \gls{tou} rate, shown in \cref{fig:charging}, features a super-off-peak rate of $\unitfrac[0.10320]{USD}{kWh}$ from $\unit[9]{am}$ to $\unit[2]{pm}$ to incentivize energy use during midday peak solar production on the California grid, and a peak rate of $\unitfrac[0.33474]{USD}{kWh}$ from $\unit[4]{pm}$ to $\unit[9]{pm}$ to disincentivize energy use during peak grid load.
As we are solving a macroscopic planning problem, we assign demand charges at the city zone level and make the demand charge continuous (instead of per $\unit[50]{kW}$ discretization) since the peak demand per zone is on the order of thousands of kilowatts.
This monthly demand charge is then appropriately rescaled for the case study time horizon (24 hours) such that $\gls{p_demand} = \unitfrac[0.056497]{USD}{kW}$.

In determining the per-vehicle fleet procurement cost, \gls{p_fleet}, we assume a depreciation rate of $\unitfrac[20]{\%}{year}$ on the original sale price.
Sale prices are provided in \cref{table:evs}. We also add an annual cost of $\unitfrac[2127]{USD}{vehicle}$ which includes the cost of insurance, registration, and finance charges \cite{aaa}.
This yearly cost is then rescaled for the case study time horizon such that $\gls{p_fleet} = 20.09, 23.12,$ and $\unitfrac[31.55]{USD}{vehicle}$ for the \gls{spring}, \gls{leaf}, and \gls{model3}, respectively.

The per-station costs for charging station installation are $\unit[2887]{USD}, \unit[5287]{USD}, \unit[28287]{USD},$ and $\unit[88187]{USD}$ for the $\unit[7.7]{kW}, \unit[16.8]{kW}, \unit[50]{kW}$, and $\unit[150]{kW}$ stations, respectively, based on Exhibit 1 of \cite{rmi}. Assuming a station lifetime of 10 years, we can compute the equivalent annual cost and then rescale for the case study time horizon to obtain $\gls{p_stn}\parentheses{\unit[7.7]{kW}} = \unitfrac[2.61]{USD}{station}$, $\gls{p_stn}\parentheses{\unit[16.8]{kW}} = \unitfrac[3.55]{USD}{station}$, $\gls{p_stn}\parentheses{\unit[50]{kW}} = \unitfrac[13.36]{USD}{station}$, and $\gls{p_stn}\parentheses{\unit[150]{kW}} = \unitfrac[41.37]{USD}{station}$. Note that this cost accounts for the procurement of charging station equipment, cables, and accompanying transformers, but does not include costs associated with land use, labor, and permitting, which vary significantly by site.

\subsection{Experimental design}
Using the model parameters outlined in \cref{sec:case_study_params}, we conduct two experiments:
\begin{enumerate}
    \item A comparison of \gls{eamod} fleet operations and station siting determined by joint optimization \cref{eq:joint_eamod_prob} between different fleet vehicle choices:  the \gls{model3} ($\unit[75]{kWh}$), the \gls{leaf} ($\unit[40]{kWh}$), and the \gls{spring} ($\unit[27.4]{kWh}$).
    \item A comparison of \gls{eamod} fleet operations and station siting determined by joint optimization \cref{eq:joint_eamod_prob} versus a ``baseline'' scenario in which an \gls{eamod} fleet optimizes operations on a fixed station siting based on present-day stations. For this comparison, the middle-sized \gls{ev}, the \gls{leaf}, is used.
\end{enumerate}

In the baseline scenario comparison, we determine the charging station siting from the US Department of Energy's \gls{afdc} API. The API returns the number of Level 2 and DC fast charging plugs for every recorded charging station in San Francisco. We assume Level 2 stations are $\unit[7.7]{kW}$ and DC fast charging stations are $\unit[50]{kW}$, which are common charging rates currently used. As \gls{ev} penetration is much lower in present day compared to our joint optimization future mobility scenario in which all travel is served by \glspl{ev}, we scale up the present-day station distribution by a constant factor of 59.86 such that the city total installed charging infrastructure capacity is equal to the total installed capacity resulting from the joint optimization, $\unit[629.91]{MW}$.

\subsection{Results and discussion}
\paragraph{Comparison between \gls{ev} models} A summary of results for the comparison between \gls{ev} models is presented in \cref{table:evs}.

\begin{table}
\centering
\caption{Comparison of specifications, total costs, and charging results between EV models of varying battery size. 
The Dacia Spring fleet has the lowest total cost due to its high efficiency and low vehicle price.}
{\footnotesize
\begin{tabular}{llccc}
\hline
 & Unit & Model 3 & Leaf S & Spring \!\!\\
\hline
Vehicle price & USD & $\num{46990}$ & $\num{31600}$ & $\num{26058}$ \\
Efficiency & Wh/km & $\num{173}$ & $\num{189}$ & $\num{119}$ \\
Battery size & kWh & $\num{75}$ & $\num{40}$ & $\num{27.4}$ \\
\hline
Fleet size & & $\num{154.77}$k & $\num{154.77}$k & $\num{154.77}$k \\
\hline
Fleet procurement & USD & $\num{4883.82}$k & $\num{3579.49}$k & $\num{3109.79}$k \\
cost &&&&\\
Charging station & USD & $\num{118.67}$k & $\num{137.15}$k & $\num{95.56}$k \\
procurement cost &&&&\\
Electricity cost, & USD & $\num{721.01}$k & $\num{774.03}$k & $\num{540.74}$k \\
(energy) &&&&\\
Electricity cost, & USD & $\num{22.30}$k & $\num{26.19}$k & $\num{18.18}$k \\
(demand charges) &&&&\\
Rebalancing cost  & USD & $\num{81.10}$k & $\num{81.20}$k & $\num{77.77}$k \\
\textbf{Total} & USD & $\num{5826.90}$k & $\num{4598.06}$k & $\num{3842.05}$k \\
\hline
Charging energy & MWh & $\num{6039.46}$ & $\num{6521.18}$ & $\num{4555.97}$ \\
consumed &&&&\\
Peak charging & MW & $\num{394.72}$ & $\num{463.51}$ & $\num{321.86}$ \\
load &&&&\\
Rebalancing & km & $\num{1749.63}$ & $\num{1751.80}$ & $\num{1677.80}$ \\
distance traveled &&&&\\
\hline
\end{tabular}
}
\label{table:evs}
\end{table}

Fleet procurement cost is the most significant cost term, accounting for $\unit[77.8]{\%}$ to $\unit[83.8]{\%}$ of total costs.
For all \gls{ev} models, the optimization prioritizes having the fewest number of vehicles possible to satisfy peak travel demand and any necessary rebalancing between trips.
As a result, the number of vehicles determined by the optimization is identical for all \gls{ev} models, with $\gls{n_fleet} = \unit[155.77]{k}$.

\begin{figure}
    \centering
    \begin{subfigure}[b]{0.95\linewidth}
        \centering
        \framebox{
        \includegraphics[width=0.85\textwidth]{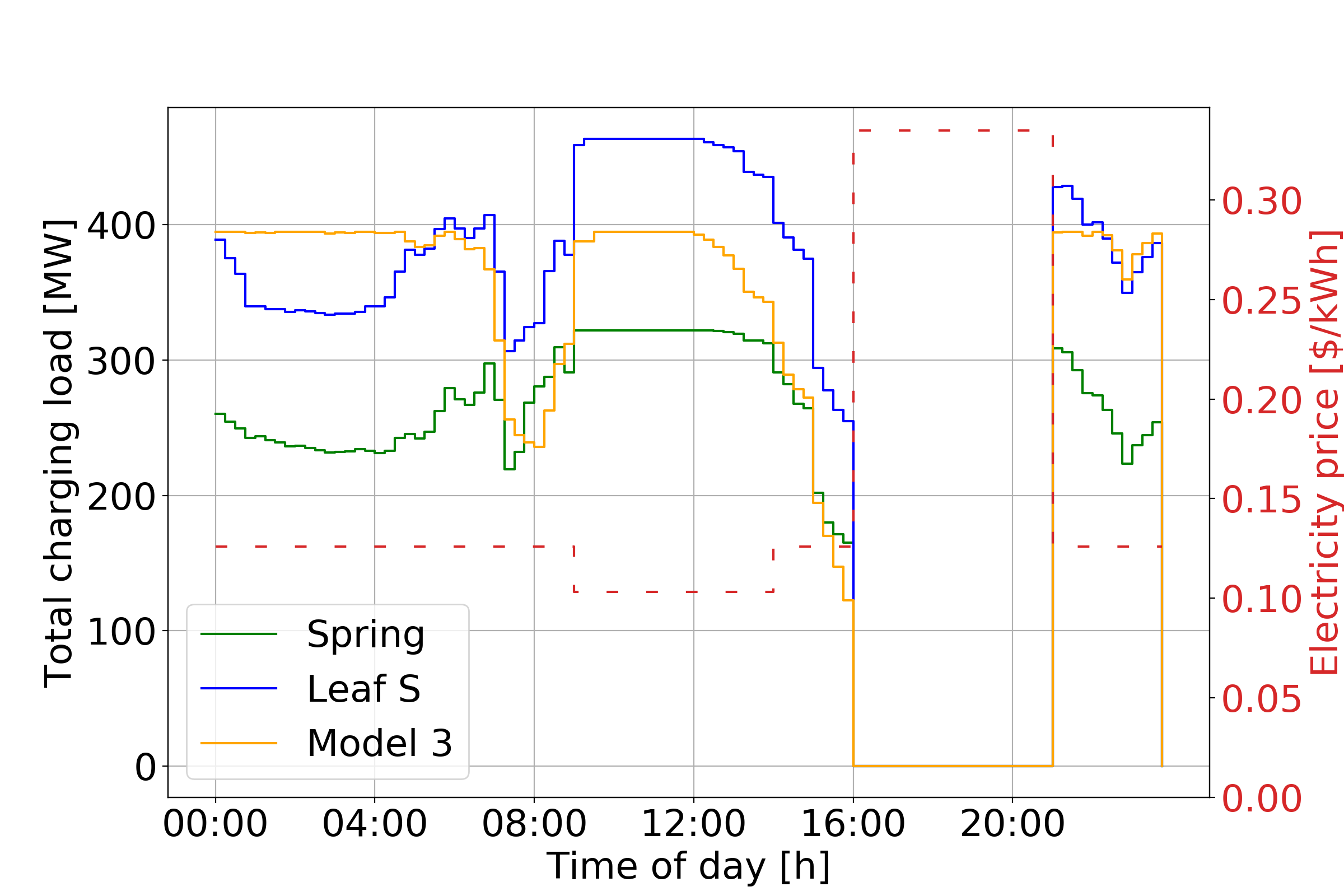}
        }
        \caption{\label{fig:charging_evs}}
    \end{subfigure}
    
    \begin{subfigure}[b]{0.95\linewidth}
        \centering
        \framebox{
        \includegraphics[width=0.85\textwidth]{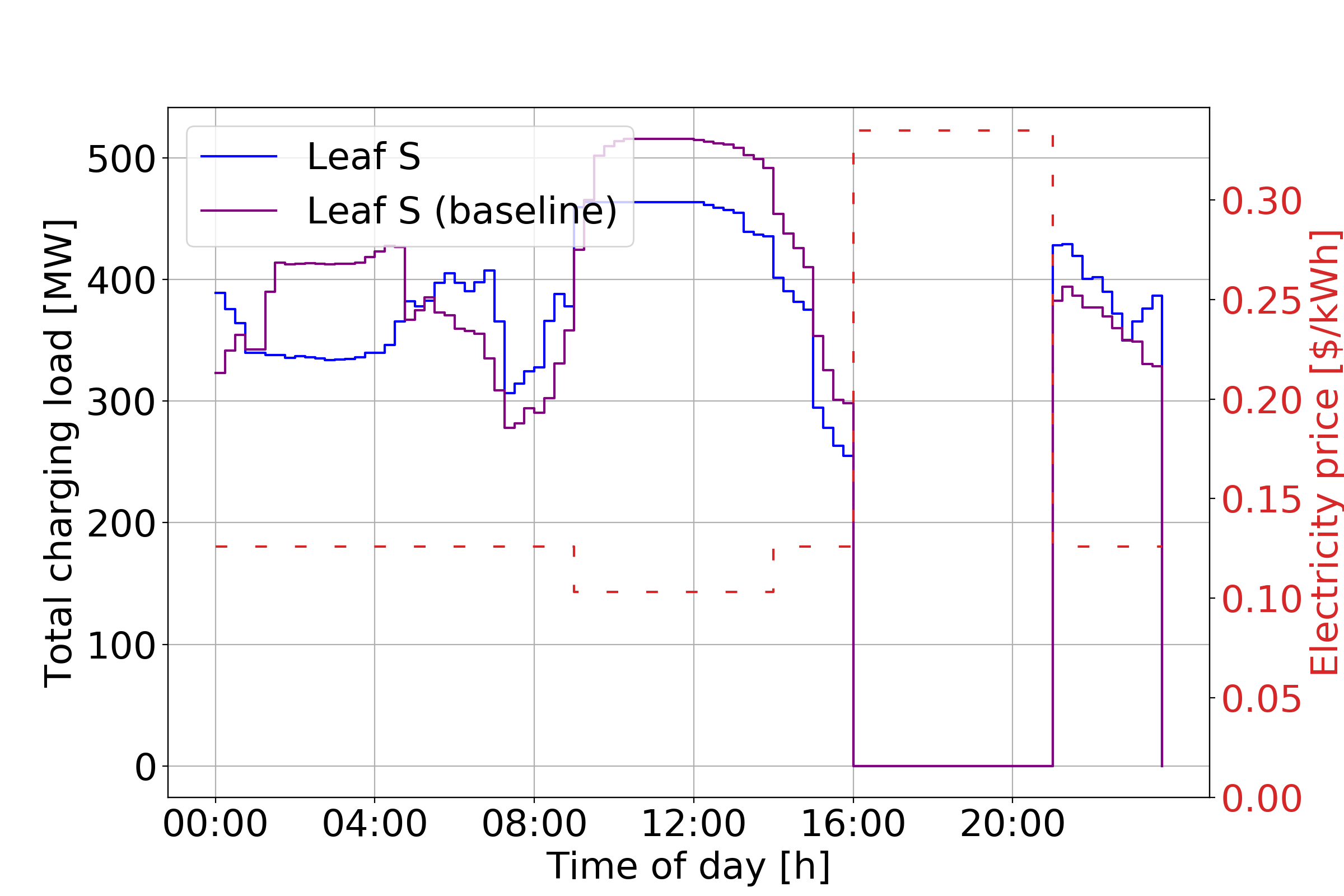}
        }
        \caption{\label{fig:charging_baseline}}
    \end{subfigure}
    \caption{\label{fig:charging}\textbf{(a)} Comparison of total charging load between fleets of varying \gls{ev} models.
    \textbf{(b)} Comparison of total charging load for Leaf S fleet between optimized station siting and baseline station siting scenario. In both figures, the \gls{tou} electricity price is plotted against the right axis.}
\end{figure}

Energy electricity cost is the second largest cost term for all three \gls{ev} models, accounting for $\unit[12.4]{\%}$ to $\unit[16.8]{\%}$ of total cost.
As shown in \cref{fig:charging}, the \gls{eamod} fleet has sufficient charge scheduling flexibility to completely avoid charging during the highest price period from $\unit[4]{pm}$ to $\unit[9]{pm}$.
Conversely, the fleet's peak charging occurs during the lowest price period from $\unit[9]{am}$ to $\unit[2]{pm}$.

\begin{figure}
	\centering
	\framebox{
	\includegraphics[width=0.85\linewidth]{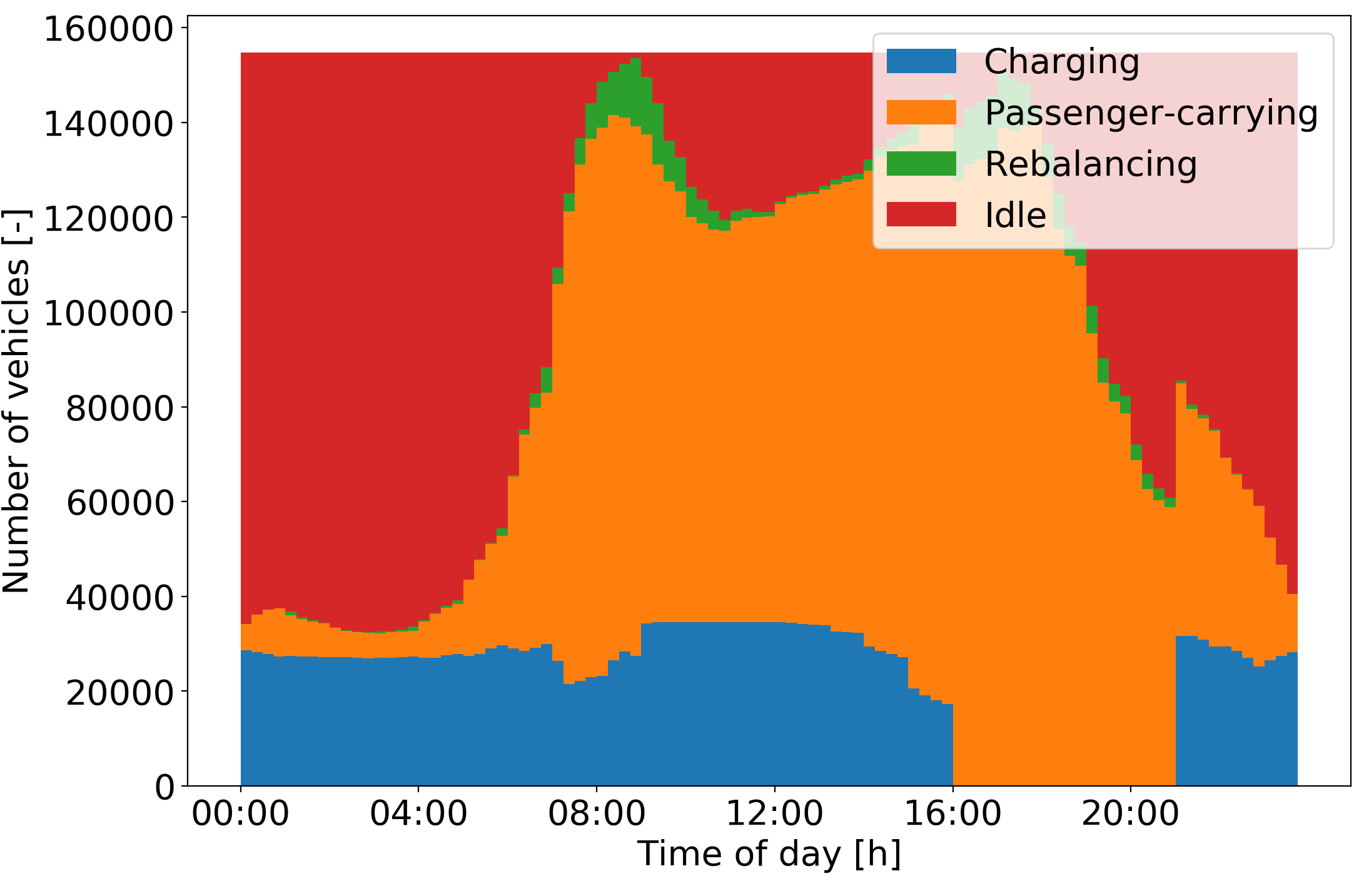}
	}
	\caption{\label{fig:vehicle_status_leafS} Distribution of vehicle status over time for the \gls{leaf} fleet, with charging station siting jointly optimized.}
\end{figure}

\begin{figure}
	\centering
	\begin{subfigure}[b]{0.95\linewidth}
	    \centering
	    \framebox{
	    \includegraphics[width=0.85\linewidth]{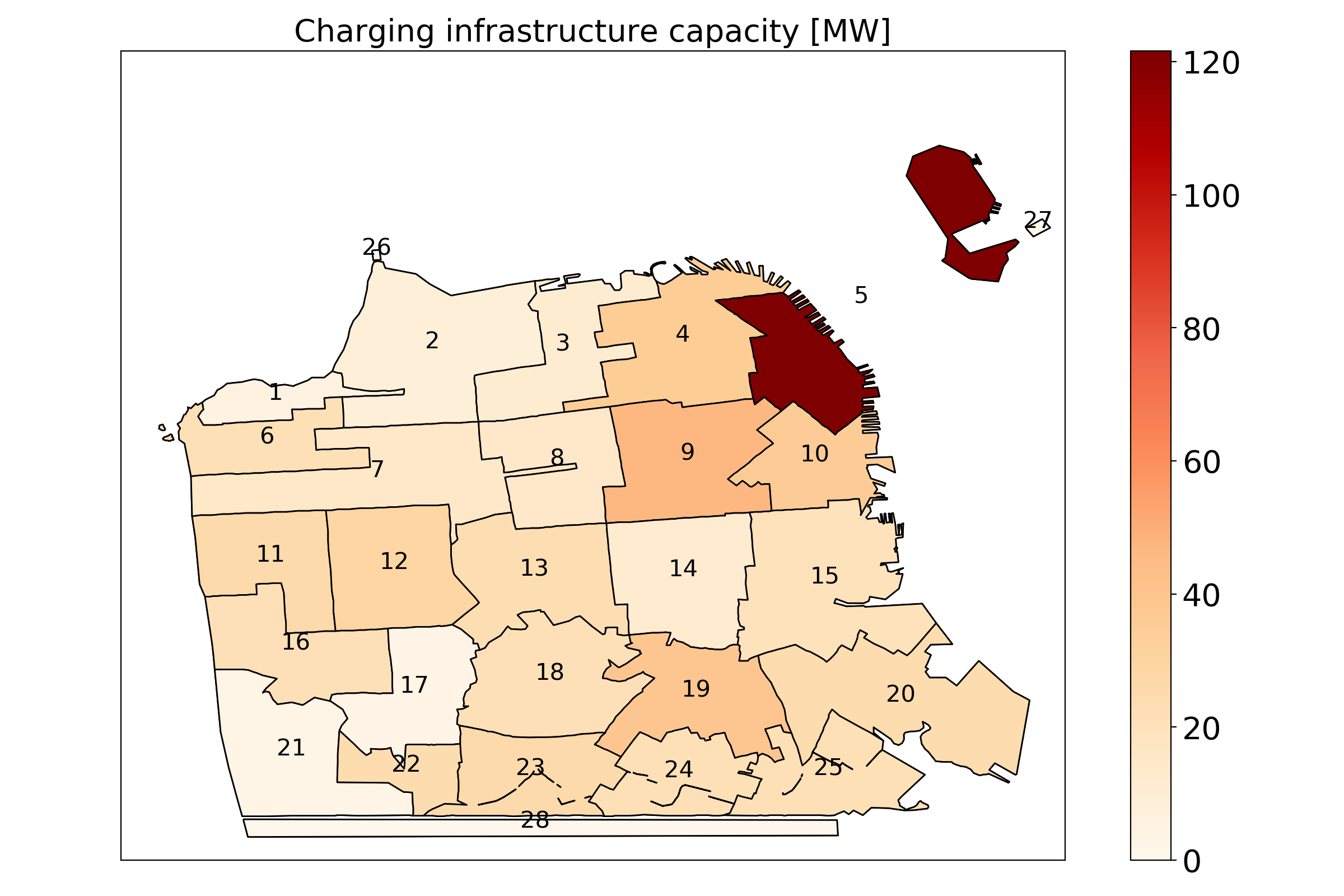}
	    }
	    \caption{\label{fig:heatmap_leafS}}
	\end{subfigure}
	
	\begin{subfigure}[b]{0.95\linewidth}
	    \centering
	    \framebox{
	    \includegraphics[width=0.85\linewidth]{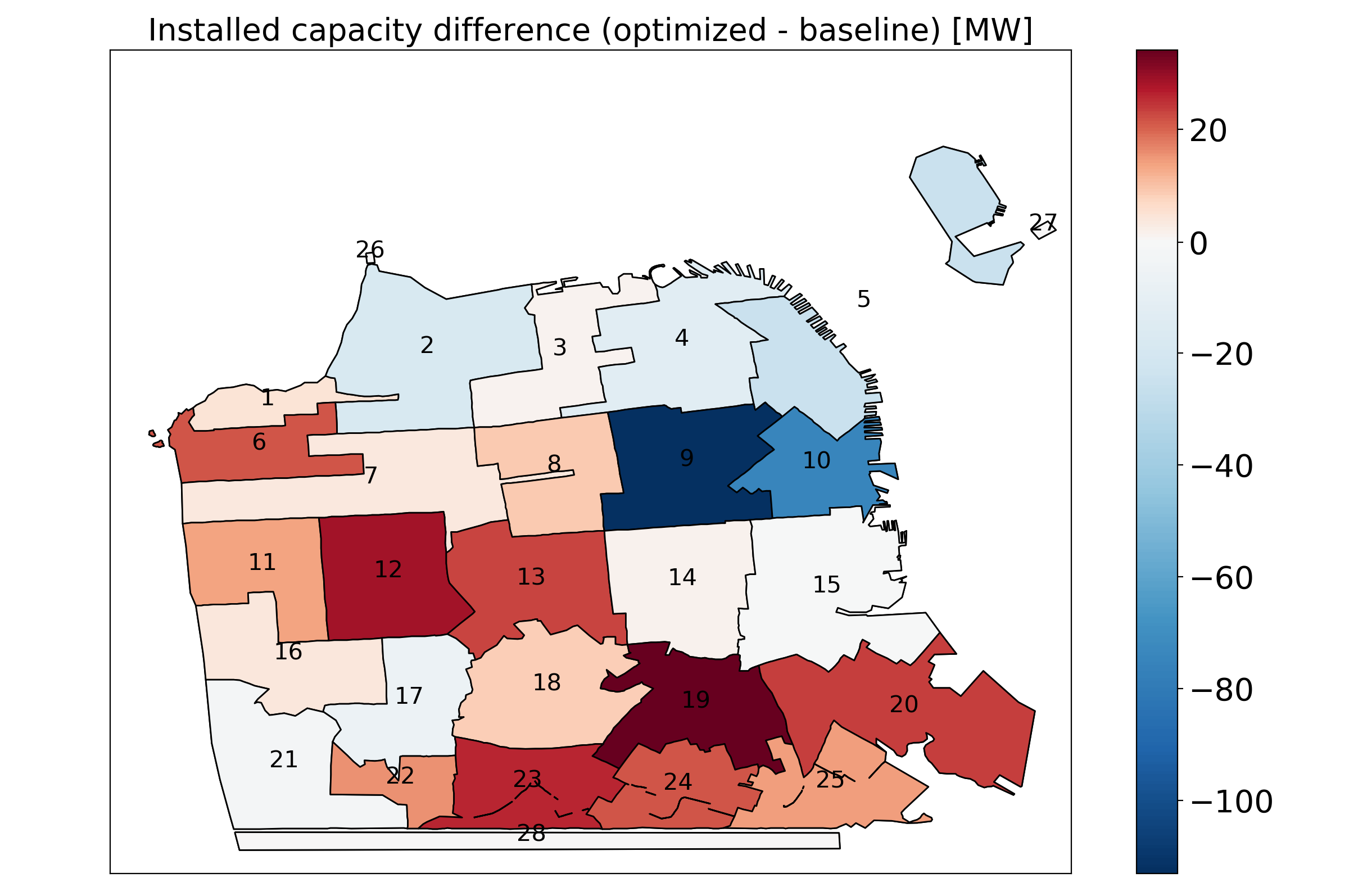}
	    }
	    \caption{\label{fig:heatmap_infra_diff}}
	\end{subfigure}
	\caption{\label{fig:heatmap}\textbf{(a)} Heat map of installed charging capacity resulting from the siting optimization with the \gls{leaf} fleet. \textbf{(b)} Heat map of the difference in installed charging capacity for a \gls{leaf} fleet between the joint charging station siting optimization scenario and the baseline siting scenario. Red indicates greater installed capacity in the optimized scenario whereas blue indicates greater capacity in the baseline scenario.}
\end{figure}

\begin{figure}
    \centering
    \framebox{
    \begin{subfigure}[b]{0.465\linewidth}
        \centering
            \includegraphics[width=\textwidth]{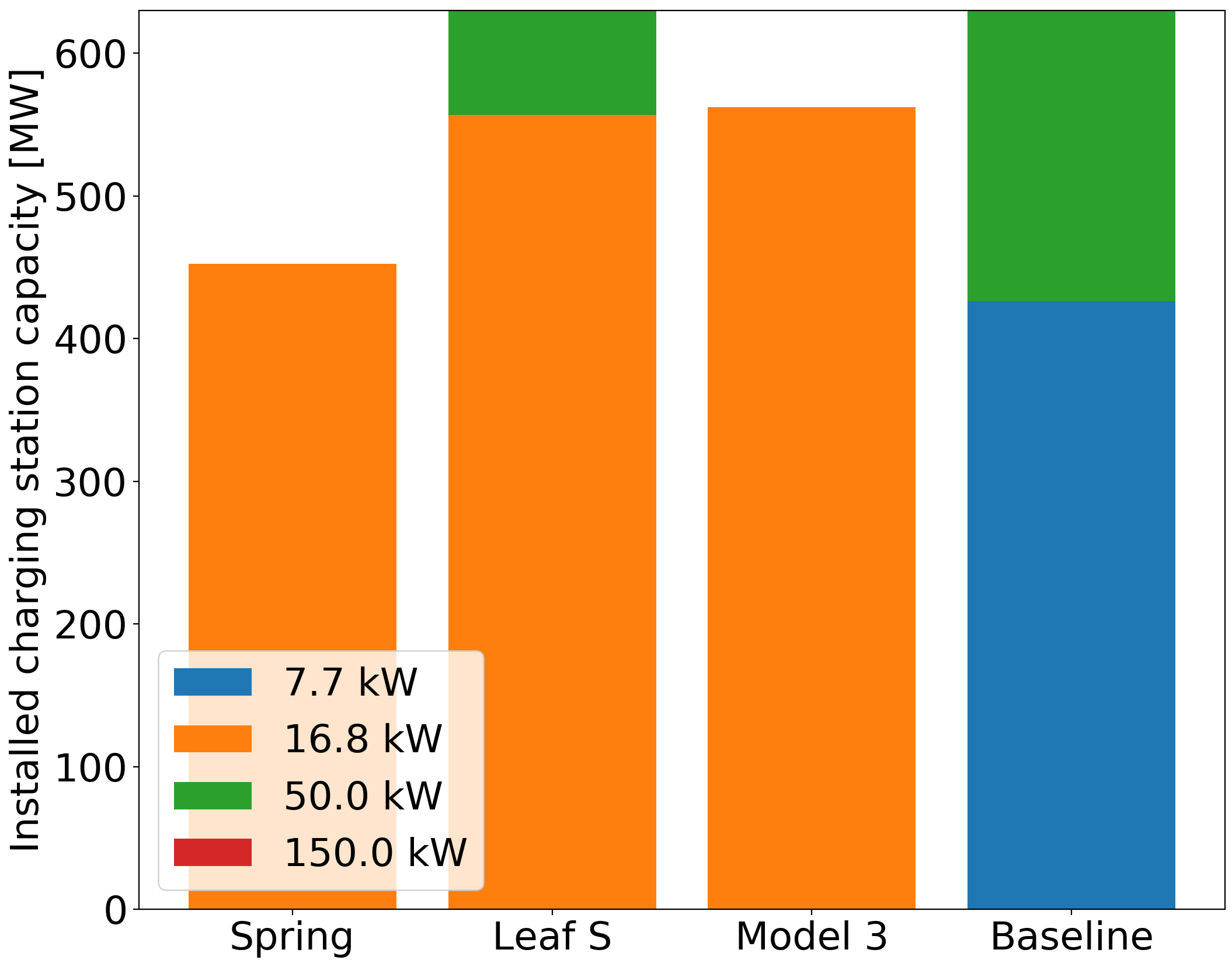}
        \caption{\label{fig:infra_cap_rating}}
    \end{subfigure}
    \begin{subfigure}[b]{0.465\linewidth}
        \centering
        \includegraphics[width=\textwidth]{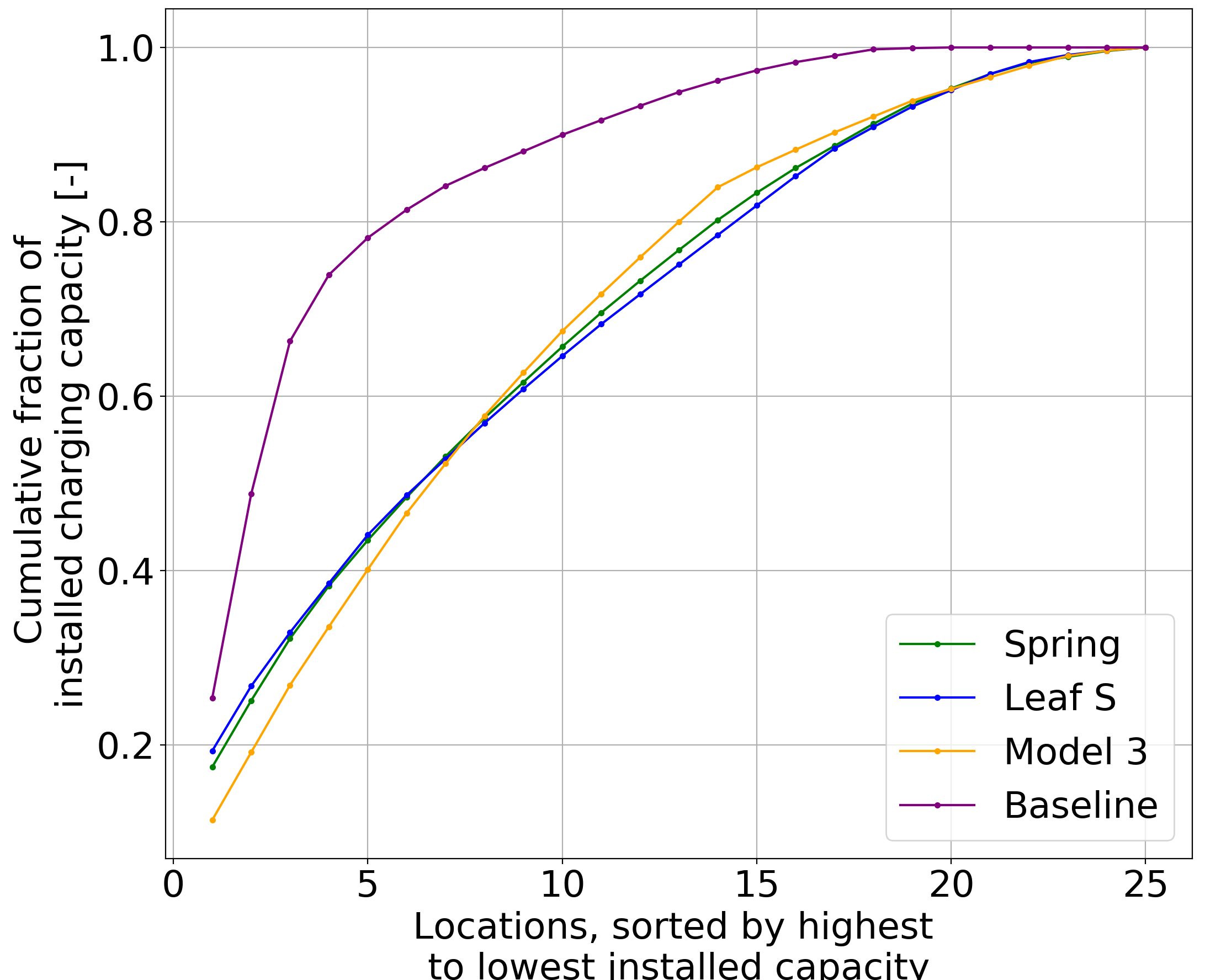}
        \caption{\label{fig:infra_cap_location}}
    \end{subfigure}
    }
    \caption{\textbf{(a)} Comparison of installed charging capacity by charging rate between fleets of varying \gls{ev} models.
    \textbf{(b)} Comparison of the cumulative distribution of installed charging capacity across locations, between the baseline siting scenario and optimized siting results using varying \gls{ev} model fleets.}
\end{figure}

The given travel demand also plays a key role in determining charge scheduling and station siting.
The macroscopic pattern in the San Francisco transportation system is for vehicles to begin the day primarily in the non-commercial zones, then concentrate in the downtown commercial district (Zones 4, 5, 9, and 10; see map in \cref{fig:heatmap}) in the morning rush hour between $\unit[7]{am}$ and $\unit[9]{am}$, then return to the non-commercial zones in the evening rush hour between $\unit[4]{pm}$ and $\unit[6]{pm}$.
The temporal aspect of this travel pattern can be seen in the passenger-carrying vehicles of \cref{fig:vehicle_status_leafS}.
As a result, there is a longer period of charging primarily in non-commercial zones from after the evening rush hour to the morning rush hour of the following day, and a shorter 8-hour period during the midday working hours between the rush hours in the commercial zones.
Given the high number of vehicles in the commercial zones in the midday, a low electricity price, and a short window in which these vehicles are concentrated there, the optimization installs a greater number of stations in the commercial zones, shown for the \gls{leaf} in \cref{fig:heatmap_leafS}.
For the \gls{leaf}, $\unit[50]{kW}$ DC fast charging stations are installed at these commercial zones to ensure vehicles can quickly recharge within this midday period before the evening rush hour.
While the \gls{model3} and \gls{spring} also have greater installed capacity in the commercial zones, no DC fast charging is installed in any of the zones for either \glspl{ev}, as seen in \cref{fig:infra_cap_rating} (however, recall that fast charging above $\unit[30]{kW}$ is not an option for the \gls{spring}).
With an $\unit[87.5]{\%}$ larger battery, the \gls{model3} does not require as much recharging at midday as the \gls{leaf} to make it through the evening rush hour (and the high electricity price period).
Despite having a smaller battery, the light-weight \gls{spring} is $\unit[37.0]{\%}$ more efficient than the \gls{leaf} and also requires less midday recharging.

The advantage of the \gls{spring}'s higher efficiency is impactful across every cost term, and furthermore, reduces load in both the power and transportation systems.
In comparison with the larger battery \gls{ev} models, the \gls{spring} consumes less energy and consequently requires less installation of charging infrastructure, has a lower peak charging load, and fewer \gls{vmt} due to less rebalancing travel needed to load balance charging demand between stations.
However, in further experiments with yet smaller \gls{ev} models (not shown in the interest of space limitations), a trade-off in benefits is observed.
If the \gls{ev} model's battery is too small, then some vehicles must be recharged during the evening rush hour, which requires a larger fleet size beyond $\gls{n_fleet} = \unit[155.77]{k}$, thereby increasing the fleet procurement cost.
Furthermore, this recharging takes place during the most expensive price period.
Further reducing the battery size, the optimization problem becomes infeasible as fleet vehicles are unable to serve the longest trips from San Francisco to the Peninsula without becoming stranded, as the charging infrastructure is only made available within the region of operation in this experimental setup.

\paragraph{Optimized vs. baseline station siting} \cref{table:baseline} compares the baseline with the optimized scenario.

\begin{table}
\centering
{\small
\begin{tabularx}{0.485\textwidth}{llccc}
\hline
 & Unit & Leaf S,  & Leaf S, & Change  \!\!\\
 & & Baseline & Optimized & \\
\hline
Fleet size & & $\num{154.77}$k & $\num{154.77}$k  & $\num{0.00}$\% \\
\hline
Charging station & USD & $\num{199.10}$k & $\num{137.15}$k & $\num{-31.11}$\% \\
procurement cost &&&&\\
Electricity cost, & USD & $\num{806.92}$k & $\num{774.03}$k & $\num{-4.08}$\% \\
energy &&&&\\
Electricity cost, & USD & $\num{29.12}$k & $\num{26.19}$k & $\num{-10.07}$\% \\
demand charges &&&&\\
Rebalancing cost  & USD & $\num{91.44}$k & $\num{81.20}$k & $\num{-11.20}$\% \\
\textbf{Total} & USD & $\num{1126.57}$k & $\num{1018.57}$k & $\num{-9.59}$\% \\
\hline
Charging energy & MWh & $\num{6819.27}$ & $\num{6521.18}$ & $\num{-4.37}$\% \\
consumed &&&&\\
Peak charging & MW & $\num{515.42}$ & $\num{463.51}$ & $\num{-10.07}$\%  \\
load &&&&\\
Rebalancing & km & $\num{1972.65}$ & $\num{1751.80}$ & $\num{-11.20}$\% \\
distance traveled &&&&\\
\hline
\end{tabularx}
}
\caption{Comparison of total costs and charging results for the Leaf S fleet between an optimized charging station scenario and the baseline scenario.}
\label{table:baseline}
\end{table}

Jointly optimizing the charging station siting with the \gls{eamod} operations lowers total cost, total vehicle travel, and peak charging load on the electric grid compared to using a baseline station siting based on the present-day siting.
Despite having more stations in commercial zones compared to non-commercial zones, the station siting resulting from the joint optimization is still more spatially distributed than present-day station siting for which the majority of capacity is concentrated in commercial zones, as shown in \cref{fig:heatmap_infra_diff}.
\cref{fig:infra_cap_location} shows that in the baseline scenario, the top four city zones by installed charging capacity ($\unit[16]{\%}$ of all zones), which are the commercial zones, already make up $\unit[74]{\%}$ of the total installed capacity.
In contrast, it requires the top 13 stations ($\unit[52]{\%}$ of all zones) in the optimized siting scenario to attain that same share of capacity.
The present-day siting is justifiable given low \gls{ev} penetration and no automation, since it captures charging business in high traffic regions at times of day when people are active.
However, in a future mobility scenario with a high penetration of \gls{eamod}, the optimal siting is spread more uniformly throughout the city.
Consequently, rebalancing \gls{vmt} is reduced by $\unit[11.20]{\%}$ as less repositioning is required for vehicles to access available charging stations, which in turn contributes to a $\unit[4.08]{\%}$ reduction in energy consumption and a $\unit[10.07]{\%}$ reduction in peak charging load, as shown in \cref{table:baseline}.
Additionally, as the fleet vehicles are autonomous and can operate at all hours of the day, they are not limited to charging during the midday when there is a large share of vehicles in the commercial zones.
The ability for vehicles to autonomously take turns recharging in the middle of the night, when they are largely in non-commercial zones, allows for greater distribution of charging throughout the day and a lower peak load in the joint optimized scenario, as shown in \cref{fig:charging_baseline}.
Notably, there is a $\unit[64]{\%}$ reduction in installed DC fast charging capacity in favor of high-power Level 2 AC charging.

The joint optimization problem \cref{eq:joint_eamod_prob} for the \gls{leaf} fleet, which has 5.3 million decision variables, was solved using Gurobi solver in 61 iterations over 2.33 hours on a compute instance with 24 vCPU and 64GB RAM.

\section{Conclusion}
\label{sec:conclusion}
This paper explored the benefits of optimizing the operations of a fleet of electric Autonomous Mobility-on-Demand (E-AMoD) vehicles jointly with the siting of the charging infrastructure.
In particular, we devised a network flow model capturing the movements and charging activities of the electric vehicles (EVs) in time, and integrated it within the static charging infrastructure siting problem.
The resulting joint design and control problem is convex and can be solved to global optimality with off-the-shelf linear programming algorithms.
Our real-world case-studies compared the total costs achievable with three different types of \glspl{ev}, revealing that the lightest and cheapest \gls{ev}, despite its limited range, would result in the lowest total cost, and that changing the vehicle type would significantly affect the resulting infrastructure design.
Finally, we quantified the benefits of jointly optimizing the siting of the chargers.
Our results revealed that, compared to the case where the infrastructure siting corresponds to a linear scale-up of the present-day layout and only the E-AMoD operations are optimized, our joint optimization framework can reduce the total cost incurred by the \gls{eamod} operator, the peak charging load, and the empty-vehicle distance traveled by up to 10\%, and also lower the charging station procurement cost by more than 30\%.
In particular, the share of charging capacity provided by DC fast charging stations is reduced nearly three-fold in favour of high-power Level 2 AC stations.

This work can be extended as follows:
First, we would like to capture the effects of \gls{ev} charging on the power grid, potentially jointly optimizing its design and enabling vehicle-to-grid operations.
Second, it is of interest to study the impact of heterogeneous fleets consisting of differently sized EVs, and of hybrid electric and internal combustion engine vehicles.
Finally, we would like to investigate the sensitivity of our results with respect to travel demand variability and different energy consumption models.


\addtolength{\textheight}{-12cm}   




\jvspace{-.5em}
\section*{Acknowledgments}
J. Luke would like to thank StreetLight Data, Inc. for providing travel demand data of San Francisco under the StreetLight Academic Access Agreement.
The authors thank Dr.\ Ilse New and Dr.\ Edward Schmerling for proofreading this paper.
This research was supported by the National Science Foundation under the CPS program and the Stanford University Bits \& Watts EV50 Project. This article solely reflects the opinions and conclusions of its authors and not NSF or Stanford University.


\jvspace{-.5em}
{\small
\printbibliography
}

\end{document}